\documentclass[a4paper]{jpconf}
\usepackage{graphicx}

\usepackage{delarray,amsmath,bbm,amssymb}
\usepackage{cite}

\newcommand{\beq} {\begin{equation}}
\newcommand{\eeq} {\end{equation}}
\newcommand{\beqa} {\begin{eqnarray}}
\newcommand{\eeqa} {\end{eqnarray}}

\newcommand{\bos}[1]{\boldsymbol{#1}}
\newcommand{\lqcd}{\Lambda_{QCD}}
\newcommand{\kvec}{{{\bos{k}}_\perp}}
\newcommand{\kt}{k_\perp}
\newcommand{\gsim}{\gtrsim}
\newcommand{\mM}{\mathcal{M}}
\newcommand{\im}{{\rm Im}}

\newcommand{\mrm}[1] {{\mathrm{#1}}}
\newcommand{\eq}[1]{(\ref{eq:#1})}
\newcommand{\ie}{{\it i.e.}}

\begin{document}
\title{Spin asymmetry at large $x_F$ and $k_\perp$}

\author{Matti J\"arvinen\footnote[1]{Present address: Institut for Fysik og Kemi, Syddansk Universitet, Campusvej 55, 5230 Odense M, Denmark}}

\address{Department of Physical Sciences and Helsinki Institute of Physics, POB 64, FIN-00014 University of Helsinki, Finland}

\ead{mjarvine@ifk.sdu.dk}

\begin{abstract}
We suggest that the large single spin asymmetries observed at high momentum fractions $x_F$ and transverse momenta $k_\perp$ of the pion in $p^\uparrow p \to \pi(x_F,k_\perp)+X$ arise from the coherence of the soft interactions with the hard parton scattering process. Such coherence can be maintained if $x_F \to 1$ as $k_\perp \to \infty$, while $k_\perp^2(1-x_F) \sim \lqcd^2$ stays fixed. Analogous coherence effects have been seen experimentally in the Drell-Yan process at high $x_F$. We find that the $p^\uparrow p \to \pi X$ production amplitudes have large dynamic phases and that helicity flip contributions are unsuppressed in this limit, giving rise to potentially large single spin asymmetries.
\end{abstract}

\section{Introduction and motivation}

Single pin asymmetry (SSA) is the dependence of a cross section on a single measured spin. The size of a transverse SSA is characterized by the analyzing power
\beq \label{eq:ANdef}
 A_N 
 =  \frac{d\sigma^\uparrow-d\sigma^\downarrow}{d\sigma^\uparrow+d\sigma^\downarrow} \propto \im[\mM_\rightarrow\mM_\leftarrow^*]
\eeq
where $\updownarrow$ ($\leftrightarrow$) refer to the transverse spin (helicity) of the polarized particle. Sizeable $A_N$'s have been observed in $p^\uparrow p \to \pi(x_F,\kvec)+X$ \cite{Adams:1991rw,Adams:2003fx} and in $pp \to \Lambda^\uparrow(x_F,\kvec)+ X$ \cite{Bunce:1976yb}. At the highest measured longitudinal momentum fractions $x_F \simeq 0.8$ of the pion in $p^\uparrow p \to \pi^\pm X$ at the E704 experiment \cite{Adams:1991rw} ($\sqrt{s}=20$~GeV) the asymmetry rises to $|A_N| \sim 0.4$, and increases for transverse momenta above $\kt = 0.7$~GeV. These asymmetries are an order of magnitude larger than those observed in DIS ($ep^\uparrow \to e+ X$) \cite{Airapetian:2004tw}. Recently the asymmetry has been seen to persist at much larger center of mass energy $\sqrt{s}=200$~GeV in $\pi^0$ production at STAR \cite{Adams:2003fx}. The asymmetry increases with $k_\perp$ up to $\kt \simeq 2.5$~GeV for all available $x_F=0.25 \ldots 0.56$.

From \eq{ANdef} we see that a sizeable $A_N$ requires a helicity flip and a large (dynamical) phase between the two helicity amplitudes. Due to these requirements $A_N$ vanishes the standard leading twist collinear QCD factorization \cite{Kane:1978nd} but can be described by using generalized schemes. The E704 and STAR data have been fitted using transverse momentum dependent factorization \cite{Anselmino:1994tv} and including twist-three effects \cite{Qiu:1998ia}. However, while these approaches are able to reproduce the $x_F$ dependence of $A_N$, they predict that $A_N \propto \lqcd/\kt$ in apparent conflict with the data.

Notice that largest asymmetries have been observed at very large $x_F \simeq 0.8$ values of pion momentum fraction in $p^\uparrow p \to \pi(x_F,\kt)+ X$. In order to produce such a pion within the standard leading twist QCD factorization one quark must carry a momentum fraction $x \gsim 0.9$ of the proton and transfer a fraction $z \gsim 0.9$ of its momentum to the pion. These stringent requirements imply a very small production cross section. In fact, QCD at leading twist was found to underestimate the E704 pion production cross section by an order of magnitude at high $x_F$ \cite{Bourrely:2003bw} which casts doubt on the applicability of factorization based approaches on SSAs in this kinematic region. On the other hand, the cross section measured at higher $\sqrt{s}$ and lower $x_F$ by STAR is consistent with leading twist QCD.

\section{Coherence at large $x_F$}

As noted above present approaches fail to produce the total cross section at large $x_F$ and the $k_\perp$ dependence of $A_N$ in $p^\uparrow p \to \pi(x_F,\kt)+  X$.
These shortcomings motivate us to suggest that the asymmetry is a large $x_F$ coherence effect \cite{Hoyer:2006hu}. Such effects in unpolarized Drell-Yan have been studied previously \cite{Berger:1979du}. The angular distribution of the muon pair in $\pi N \to \mu\mu(x_F)+ X$ provides a clear signature of the coherence effects which set in at high $x_F$. When the intrinsic hardness of the contributing pion Fock states becomes comparable to the virtuality $Q^2$ of the photon the angular distribution of the muons, which is $1+\cos^2\theta$ at leading twist, turns into $\sin^2\theta$. This phenomenon was subsequently observed in the Drell-Yan data \cite{Anderson:1979xx} where the change of angular distribution occurs at $x_F \simeq 0.7$ for $Q^2 \simeq 20$~GeV$^2$.

In general, the increase of coherence effects at large $x_F$ can be understood as follows (see \cite{Brodsky:1991dj}). The lifetime $\tau$ of a Fock state inside a rapidly moving proton is the inverse of the (light-front) energy difference $\Delta E$ between the Fock state and the proton
\beq \label{eq:DEexpr}
 P^+\Delta E = M^2 - \sum_i\frac{k_{i\perp}^2+m_i^2}{x_i}\mrm{;}\qquad \sum_ix_i=1
\eeq
where $x_i(>0)$, $k_{i\perp}$, and $m_i$ are the momentum fraction, the transverse momentum, and the mass of parton $i$, respectively, 
$P^+$ is the proton light-front momentum, and $M$ is the proton mass. When one quark carries a large $x_i\sim x_F \to 1$ all other partons 
must have $x_j \sim 1-x_F \to 0$. Hence the lifetime of the state $\tau \sim 1/\Delta E \sim (1-x_F)P^+/\lqcd^2$ becomes short. The incoherence of such state with a hard quark requires $(1-x_F)P^+/\lqcd^2\gg\tau_\mrm{hard} \sim P^+/\kt^2$ or $\kt^2(1-x_F) \gg \lqcd^2$. When $x_F$ grows large enough, \ie, 
\beq \label{eq:cohcond}
 \kt^2(1-x_F) \sim \lqcd^2 \ , 
\eeq
the hard scattering becomes coherent with the soft physics. 

\section{Coherence effects in $p^\uparrow p \to \pi X$}

Sizeable coherence effects were observed in Drell-Yan for $x_F \simeq 0.8$ and for much larger photon virtuality $Q \simeq 4-5$~GeV than the typical $\kt \sim 1$~GeV of the pion at E704. Hence coherence is expected to be significant in the kinematic region of E704 where the asymmetries are large. We study the  coherence effects in  $p^\uparrow p \to \pi X$ in a similar manner as in Drell-Yan above. We take the scale $\sim \kt$ of perturbative QCD to be large,  $\kt \to \infty$, but we keep $\kt^2(1-x_F)$ fixed (instead of $x_F$) so that \eq{cohcond} holds. In this limit single quark factorization fails and several partons from the same parent hadron contribute coherently to the process. 

\begin{figure}[h]
\includegraphics[width=0.65\textwidth]{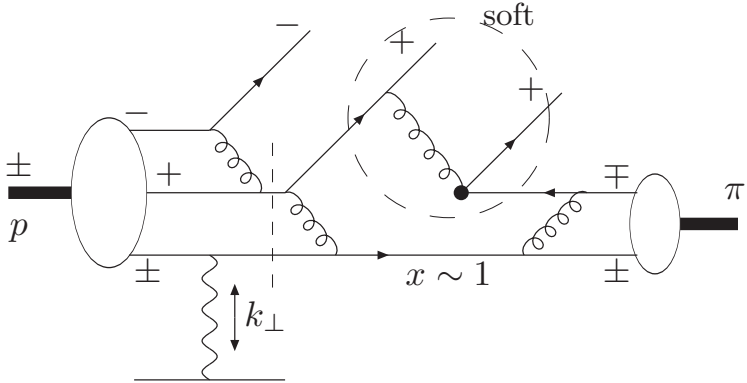}\hspace{1pc}%
\begin{minipage}[b]{12pc}\caption{\label{fig:scenario}Our mechanism for the SSA in $p^\uparrow p \to \pi X$. See text for explanation.}
\end{minipage}
\end{figure}

A leading contribution to the process in the limit of fixed $\kt^2(1-x_F)$ is shown in figure~1. A short lived Fock state with one fast quark ($x \sim 1$) is created via gluon exchange. The fast quark scatters with the target obtaining a large transverse momentum $\kt$. The quark finally picks up a slow antiquark and the pion is formed through a gluon exchange which equalizes the momentum fractions. The interactions within the slow quark system (indicated by the dashed circle in figure~1) are soft with transverse momentum scale $\sim \lqcd$. The condition \eq{cohcond} implies that all  parts of the diagram are fully coherent. 

Recall that we need a helicity flip and a large, helicity-dependent phase to create a sizeable asymmetry. As a helicity flip is suppressed in the hard interactions, the flip must occur in the soft subprocesses. This is modeled in figure~1 by helicity changing one gluon exchange. The helicity flip vertex is indicated by a dot and $\pm$ are the helicities of the quarks in the two interfering amplitudes. A dynamical phase is obtained from the hard subprocess as indicated by the vertical dashed cut.

Using some further simplifications the two helicity amplitudes of figure~1 can be estimated \cite{Hoyer:2006hu}. A sizeable helicity dependent phase indeed arises when the coherence condition \eq{cohcond} is satisfied. This is a proof of principle that large $A_N$ is possible in the kinematic limit of fixed $\kt^2(1-x_F)$.

\section{Conclusion}

We demonstrated that large $x_F$ coherence effects  may explain the large asymmetries of $p^\uparrow p \to \pi(x_F,\kt)+ X$. In the limit of large $\kt$ with $\kt^2(1-x_F)$ fixed the overall coherence of the scattering process is maintained which leads to large dynamical phases and possibly to large single spin asymmetries. Our mechanism may be able to reproduce the experimental result that $A_N$ increases with $k_\perp$ even for $k_\perp \gsim \lqcd$. Note that simply assuming that the maximum of $A_N$ as a function of $\kt$ is set by the coherence requirement \eq{cohcond} shifts the maximum from the standard expectation $k_{\perp,\mrm{max}} \sim \lqcd$ to a larger value.

\ack This work was supported in part by the Academy of Finland through grant 102046, by GRASPANP, the Finnish Graduate School in Particle and Nuclear Physics, and by the Magnus Ehrnrooth foundation.

\section*{References}
\providecommand{\newblock}{}

\end{document}